\documentclass{fundam}

\usepackage[ruled]{algorithm2e}

\usepackage{multirow}
\usepackage{amsmath,amssymb,amsfonts}
\usepackage{pstricks}
\usepackage{pst-plot}
\usepackage{subfigure}
\usepackage{algorithmic}
\usepackage{multirow}
\usepackage{mathtools}
\usepackage{url}
\usepackage{tabularx,booktabs}

\usepackage{pgfplots}
\pgfplotsset{compat=1.12}

\newcounter{myexamplecounter}
\newtheorem{property}{Property}

\newtheorem{statement}{Statement}

\begin{document}

\setcounter{page}{41}
\publyear{22}
\papernumber{2141}
\volume{188}
\issue{1}

   \finalVersionForARXIV

  \title{Universal Address Sequence Generator for Memory \\
 Built-in Self-test\thanks{This work was supported by the grant WZ/WI-IIT/2/2020 from Bialystok University of Technology,
                  Faculty of Computer Science } }

  \author{Ireneusz Mrozek\thanks{Address for correspondence:  Bialystok University of Technology,  Bialystok, Poland. \newline \newline
                    \vspace*{-6mm}{\scriptsize{Received July 2021; \ accepted September  2022.}}}
  \\
 Bialystok University of Technology,   Bialystok, Poland \\
i.mrozek@pb.edu.pl
\and
Nikolai A.  Shevchenko \\
Gymnasium, Darmstadt, Germany\\
nik.sh.de@gmail.com
\and Vyacheslav N. Yarmolik\\
Belarusian State University of Informatics and \\
Radioelectronics,  Minsk, Belarus \\
yarmolik10ru@yahoo.com
}

\maketitle

\runninghead{I. Mrozek et all.}{Universal Address Sequence Generator for Memory Built-in Self-test}

\vspace*{-6mm}
\begin{abstract}
This paper presents the universal address sequence generator (UASG) for memory built-in-self-test. The studies are based on the proposed universal method for generating address sequences with the desired properties for multirun march memory tests. As a mathematical model, a modification of the recursive relation for quasi-random sequence generation is used. For this model, a structural diagram of the hardware implementation is given, of which the basis is a storage device for storing so-called direction numbers of the generation matrix. The form of the generation matrix determines the basic properties of the generated address sequences. The proposed UASG generates a wide spectrum of different address sequences, including the standard ones, such as linear, address complement, gray code, worst-case gate delay, $2^i$, next address, and pseudorandom. Examples of the use of the proposed methods are considered. The result of the practical implementation of the UASG is presented, and the main characteristics are evaluated.
\end{abstract}


\section{Introduction}

The percentage of embedded memories in a chip is increasing. Thus, memory is a major portion of the current system-on-a-chip designs (occupying more than 70\%) and is expected to rise to 95\% of the area overhead ~\cite{report_ITRS_2015,Bhunia_2019, Marinissen_2005_722_727,Bushnell_2002_690}. The density of modern memory is rapidly increasing compared to random logic. Additionally, the smaller feature size and increasing space occupied by the memory on a chip result in an enormous critical chip area that may potentially have defects.

Memory faults can be divided on the basis of the number of memory cells being faulty, namely into one-cell faults (e.g., stuck-at faults, stuck open faults, and transition faults), and multiple cells faults (first of all pattern sensitive faults - PSF). The ﬁrst group of faults is well detectable by existing classical march tests. In the case of the second group of faults, the problem is much more difﬁcult.Although many approaches have been proposed in the literature \cite{Hayes_1975_150_157,Goor_1991_book,Cockburn_1995_117_122,Franklin_1996_1081_1087, Cheng_2000_401_404,  Bernardi_2010_1_10, Sfikas_2016_2339_2345,Cascaval_2004_227_243,Husum_2012_81_86,Wunderlich_1988_236_244}, the issue of efﬁcient detection of multiple cell faults is still open.

The traditional approach based on direct memory access for testing is costly in terms of silicon area, routing complexity, and test application time~\cite{Du_2005_1173}. The memory built-in self-test (MBIST) has become an attractive alternative and can offer benefits, such as at-speed testing and, therefore, high fault coverage~\cite{Aswin_2019_153_160, Harutyunyan_2019_562_575}. Traditionally, the MBIST solutions are based on march test algorithms.

Due to the linear complexity, regularity, symmetry and simplicity of the hardware implementation, the march tests are usually a preferred and often the only reasonable method for RAM testing. March test algorithms consists of a set of march elements. March elements are a finite sequence of read and write operations applied to every cell in the memory by accessing all memory addresses in any order. The order of memory cells can be the same for all march elements or can be reversed for some march elements ~\cite{Goor_1991_book, Goor_1996_420_427}.

Well known property of march tests is that for one run memory test execution there is no any special requirements for the address order, as well as for memory background~\cite{Goor_1991_book}. For any address order and memory background the number of detectable memory faults, including multi-cell faults,  will be the same and can be calculated according to the memory test detection ability.  One of the constructive solutions for achieving high fault coverage especially for complex faults, as has been shown in ~\cite{Cockburn_1995_117_122, Yarmolik_1998_354_357}, is multi-run testing. The idea of multi-run tests was originally formulated in the context of transparent testing~\cite{Nicolaidis_1992_598_607}, and later exhaustive and pseudo-exhaustive RAM testing~\cite{Karpovsky_1995_126_132, Kuhn_2006_153_158,Wagner_1987_332_343}. According to this idea, the same testing procedure is executed several times, each time with different initial conditions. As pointed out in many research studies~\cite{Nicolaidis_1992_598_607,Karpovsky_1995_126_132,Das_1997_215_229,Mrozek_2012_301_315,Yarmolik_S_V_2009_270}, transparent tests are able to cover a wide range of memory faults (theoretically all faults). In this case, the test process requires multiple runs of one or more memory tests. It is obvious that the fault coverage of such testing processes depends both on the test used (including the number of its iterations) and the memory background and/or address order in each iteration of the test~\cite{Yarmolik_S_V_2009_270}.

So, one of the key element of multi-run tests are the address sequences.  As has been mentioned for one-run memory test execution, there is no any specific requirements for the address order~\cite{Goor_1991_book}. The only restriction is that an entire set of all possible addresses has to be generated in an arbitrary order in an up and down direction. That is why a simple binary counter with an increment and a decrement by 1 mode can be used. It is another story in the case of multi-run memory tests. The high efficiency of such type of memory testing is obtained due to the detection of additional portions of the complex memory faults. Any new run of the same memory test has to be done with the new initial conditions. Usually, this can be a new memory background or address order, or both background  and address order. In this case, it is quite important to choose an appropriate set of memory addresses. For example, for two-run  memory test, we have to select two different address sequences with a different address order. There is no doubt that a different subset can result in different fault coverage.

From the above perspective the key element of the MBIST is the address sequence generator (ASG), which is the most critical part of the area of MBIST implementation. The ASG designs are very different, and the area required for the ASG varies between 26\% and 33\% of the MBIST~\cite{Goor_2011_1_6}. To detect complex and speed-related faults, the functionality of ASG should be extended and flexible ~\cite{Goor_2011_1_6, Kumar_2015_16797_16813}. The ASG must generate an appropriate set of address sequences (ASs), with the desired address switching activity.

Several MBIST architectures have been proposed in the literature~\cite{Harutyunyan_2019_562_575,Kumar_2015_16797_16813,Cascaval_2009_440_454,Goor_2010_382_387,Mukherjee_2010_441_453,Saravanan_2019_239_247,Sosnowski_2001_104_109}. In ~\cite{Kumar_2015_16797_16813, Saravanan_2019_239_247}, attempts were put forward for proposed architectures of address generators with a low transition. It has been proven that efficient employment of the ASG architecture has considerably reduced the switching activity of MBIST~\cite{Saravanan_2019_239_247}. The proposed approaches are based on a modified linear feedback shift register (LFSR), which generates the restricted sets of ASs.

In~\cite{Pavani_2016_1484_1488}, the MBIST address generator is used to implement addresses with a significantly low area, less power, and high speed based on a set of multiplexers and counters. In this paper, a new architecture is analyzed and proposed with more advantageous properties. The implementation aspects of several ASs, including {\em linear}, {\em address complement}, and {\em gray code} sequences have been analyzed. The proposed investigation supports several designs for an ASG to generate only one AS, and their combination requires additional area overhead.

To reduce the MBIST power consumption to test the memory core of a system on a chip, the design proposed in~\cite{Singh_2011)561_565} concentrated on just three types of ASG, namely LFSR-based, {\em linear}, and {\em gray code} ASs. A comparison with the standard solutions in terms of the area overhead and consumed power was presented and analyzed. The same power-reducing issue was investigated in [15] for only one LFSR-based AS.

In~\cite{Goor_2011_1_6}, the authors stated that the set of counting methods, commonly used in industry to detect different fault classes, including speed-related faults, consists of the {\em linear}, {\em address complement}, {\em gray code}, {\em worst-case gate delay}, $2^i$, {\em next address}, and {\em pseudorandom} ASs. All efforts within these investigations have been concentrated on the optimization of ASG implementation. The AS properties and implementation aspects of several ASs have been considered. As a result, a novel, systematic, high-speed, low-power, and low-overhead implementation based on an up-counter and a set of multiplexors was presented~\cite{Goor_2011_1_6}.

The complexity of the MBIST is a major design issue because it requires a large area and limits the speed of the MBIST. In addition, the restriction on the set of ASs may reduce the efficiency of the memory-test procedures. To overcome this tradeoff, in this paper, the universal address sequence generator (UASG) is proposed and analyzed. The motivation of this work is to design an efficient universal MBIST ASG that generates sufficiently different ASs, including the standard ASs, compared with known solutions. The area overhead and speed issues, as crucial aspects of ASG implementations, are compared with the existing techniques.

\section{Standard address sequences}

The address sequence is  a binary number system that satisfies the following property:
\begin{property}
	\label{prop_address_sequence_main_property}
	A binary number system  $A={a_m}{a_{m-1}} \ldots a_3a_2a_1$     consists of all possible $2^m$ binary combinations ${a_m}{a_{m-1}} \ldots a_3a_2a_1$ formed in an arbitrary order, where   $a_i \in \{0,1\}$.    Moreover, all combinations ${a_m}{a_{m-1}} \ldots a_3a_2a_1$ occur in  $A$ only once.
\end{property}

It should be noted that there is the strong requirement to generate all addresses in an arbitrary order and the same sequence of addresses in an inverse order for memory test implementation.

\medskip
Taking into account the Property~\ref{prop_address_sequence_main_property}, the address sequence $A$ consists of $N=2^m$ $m$-bit words as well as of $m$ $N$-bit sequences\index{bit sequence} $a_ma_{m-1}...a_1$.  The classical address sequence (counter address sequence)  for $m=3$ is presented in Table~\ref{tb_classical_address_sequence_m_3}.

\begin{table}[htb]
\vspace*{-2mm}
	\caption{The classical address sequence for $m=3$}
	\label{tb_classical_address_sequence_m_3}
	\center
\begin{tabular}{l|lll}
\hline
\multirow{2}{*}{Address} & \multicolumn{3}{c}{$m=3$}                             \\ \cline{2-4}
                  & \multicolumn{1}{l|}{$a_3$} & \multicolumn{1}{l|}{$a_2$} & $a_1$ \\ \hline
        0          & \multicolumn{1}{l|}{0} & \multicolumn{1}{l|}{0} &  0\\ \hline
        1          & \multicolumn{1}{l|}{0} & \multicolumn{1}{l|}{0} &  1\\ \hline
        2          & \multicolumn{1}{l|}{0} & \multicolumn{1}{l|}{1} &  0\\ \hline
        3          & \multicolumn{1}{l|}{0} & \multicolumn{1}{l|}{1} &  1\\ \hline
        4          & \multicolumn{1}{l|}{1} & \multicolumn{1}{l|}{0} &  0\\ \hline
        5          & \multicolumn{1}{l|}{1} & \multicolumn{1}{l|}{0} &  1\\ \hline
        6          & \multicolumn{1}{l|}{1} & \multicolumn{1}{l|}{1} &  0\\ \hline
        7          & \multicolumn{1}{l|}{1} & \multicolumn{1}{l|}{1} &  1\\ \hline
\end{tabular}\vspace*{1mm}
\end{table}

In the example (Table~\ref{tb_classical_address_sequence_m_3}), we have eight addresses, namely $000$, $001$, $010$, $011$,$100$, $101$, $110$, $111$, and three bit sequences $a_3=00001111$, $a_2=00110011$ i $a_1=01010101$.

Now, let us formulate the general properties of the bit sequences $a_i$ for any address sequences $A$. Let's start with the general property of $A={a_m}{a_{m-1}} \ldots a_3a_2a_1$ where $a_i \in \{0,1\}$

\begin{property}
	\label{prop_for_any_bit_sequence_there_are_the_same_zero_one}
	For any bit sequence  $a_i$ of an address sequence  $A$ there exist $2^{m-1}$ distinct binary combinations      for $a_{m}a_{m-1}\ldots a_{i+1}a_{i-1}\ldots a_3a_2a_1$ with $a_i=0$  and exactly the same number of combinations
	$a_{m}a_{m-1}\ldots a_{i+1}a_{i-1}\ldots a_3a_2a_1$ with $a_i=1$.
\end{property}

\noindent The Property~\ref{prop_for_any_bit_sequence_there_are_the_same_zero_one} allows to make the conclusion that there are  $2^{m-1}$ '0' values and the same number $2^{m-1}$ '1' values for any bit sequence $a_i$ within any binary numerical system $A={a_m}{a_{m-1}} \ldots a_3a_2a_1$.

\begin{property}
	For any two-bit sequence of bits $a_i$  and $a_j$ $i\ne j$ of an address sequence   $A$ there are exactly  $2^{m-2}$ all binary combinations, namely $00$, $01$, $10$, $11$ within the address sequence $A$.
\end{property}
\noindent The last property can be formulated for the general case as the following property.

\begin{property}
	\label{prop_for_any_bit_sequences_there_are_the_same_zero_one_general}
	For any number of bits  $r<m$ $a_i, a_j, \ldots , a_q$  within the  standard address sequence $A$, where $i \ne j \ne \ldots \ne q$, there are exactly $2^{m-r}$ all binary combinations, namely $00\ldots0, 00\ldots1, \ldots\,$, $11\ldots1$ within the address codes ${a_m}{a_{m-1}} \ldots a_3a_2a_1$.
\end{property}

For the given memory with $2^m$ cells, there is only one counter sequence described via classical algorithm. To increase the number of sequences with an entire set of $m-bit$ addresses many standard solutions have been presented~\cite{Saravanan_2019_239_247,Pavani_2016_1484_1488,Singh_2011)561_565,Yarmolik_2013_242_247,Savage_996_605_629,Yarmolik_2007_688_698,Yarmolik_S_V_2006_572_576, Chen_2007_562_568}.

Let us analyze Address bit permutation method as an example of one of AS generating methods.

\medskip
For a one $m$-bit address sequence $A$, there are $m!$ sequences of addresses as a result of bit permutation. For example, in the case of a counter sequence $A=a_2a_1$ we have only $2!=2$ sequences, but for $A=a_3a_2a_1$ with $m=3$ we can get $3!=6$ sequences. All the mentioned sequences for $m=2,3$ are shown in Table~\ref{tb_bit_permutation_sequences_of_address_m_2_m_3}.
\begin{table}[htb]
	\caption{Sequences of address for $m=2$ and $m=3$}
	\label{tb_bit_permutation_sequences_of_address_m_2_m_3}
	\center
	\footnotesize
	\begin{tabular}{c|c|c|c|c|c|c|c}
		\hline
		\multicolumn{2}{c|}{$m=2$} & \multicolumn{6}{c}{$m=3$} \\
		\hline
		$A\#1$ & $A\#2$ & $A\#1$ & $A\#2$ & $A\#3$ &  $A\#4$ & $A\#5$ &  $A\#6$ \\
		$a_2a_1$ & $a_1a_2$ & $a_3a_2a_1$ & $a_3a_1a_2$ & $a_2a_3a_1$ & $a_2a_1a_3$ & $a_1a_3a_2$ & $a_1a_2a_3$ \\
		\hline\hline
		\multirow{2}{*}{00} & \multirow{2}{*}{00} & 000 & 000 & 000 & 000 & 000 & 000\\
		\cline{3-8}
		& &  001 & 010 & 001 & 010 & 100 & 100\\
		\hline
		\multirow{2}{*}{01} & \multirow{2}{*}{10} & 010 & 001 & 100 & 100 & 001 & 010\\
		\cline{3-8}
		& &  011 & 011 & 101 & 110 & 101 & 110\\
		\hline
		\multirow{2}{*}{10} & \multirow{2}{*}{01} & 100 & 100 & 010 & 001 & 010 & 001\\
		\cline{3-8}
		& &  101 & 110 & 011 & 011 & 110 & 101\\
		\hline
		\multirow{2}{*}{11} & \multirow{2}{*}{11} & 110 & 101 & 110 & 101 & 011 & 011\\
		\cline{3-8}
		& &  111 & 111 & 111 & 111 & 111 & 111\\
		\hline
	\end{tabular}
\end{table}

\medskip
The key parameter for predicting the number $m!$ of memory address sequences depends only on the memory width $m$. For a large $m$, the value $m!$ can be approximated by Stirling'{s} approximation:
\begin{equation}
r! \approx r^r e^{-r}\sqrt{2 \pi r}.
\end{equation}
In reality, it is a large number.

\medskip
Let us sum up the above approach in terms of its implementation~\cite{Yarmolik_2005_413_418,Mrozek_2009_263}:
\begin{enumerate}
	\item For real memory, this approach allows getting enormous amounts of address sequences.
	
	\item There is substantial hardware overhead. For practical implementation, we need to use $m$ $m$-input multiplexers and $m$ $m$-bit registers to fix one of the address sequences out of all those possible.
	
	\item Decreasing in the performance in terms of delay due to multiplexing of address bits.
\end{enumerate}

\section{Proposed method}

The basic idea behind the proposed UASG is the significant expansion of the set of different ASs (with moderate hardware overhead), including the standard well-known and extensively used AS generated by the UASG. To achieve the goal, the fundamental bases of the binary vectors field are used~\cite{Boyd_2018_book}.

\medskip
The AS $A(n)=a_m(n)$ \allowbreak $a_{m-1}(n) \allowbreak  a_{m-2}(n)  \ldots \allowbreak a_2(n) a_1(n)$, where $a_i(n) \in \{0,1\}$, $i\in \{1, 2, 3, \ldots\,$, $m\}$, the $m$-dimensional binary vectors in binary space, are considered. Then, the problem of generating the desired AS can be regarded as $m$-dimensional binary vectors in binary space generation. The vector space consists of a set of elements $a_i(n)$ over which the binary addition operation, denoted by the XOR~($\oplus$) operation, is defined. The binary multiplication operation, denoted by the AND~($\times$) operation, is defined between an element $a_i(n)$ of the field and the vectors of the space. The set of linearly independent binary vectors, $v_i = \beta_1(i) \beta_2(i) \ldots \beta_{m-1}(i) \beta_m(i)$, $i=\overline{1,m} $, generates $m$-dimensional binary vectors $A(n)$, which is called a basis of the $m$-dimensional binary vector space. The set of linearly independent vectors, $v_i = \beta_1(i) \beta_2(i) \ldots \beta_{m-1}(i) \beta_m(i)$,   generates $m$-dimensional binary vectors $A(n)$ with all linear combinations:
\begin{equation}
\label{eqn_1}
A(n) =b_1(n) \times v_1   \oplus b_2(n) \times v_2   \oplus \ldots \oplus  b_m(n) \times v_m,
\end{equation}
where  $B(n) = b_m(n) b_{m-1}(n) b_{m-2}(n) \ldots b_2(n) b_1(n)$; $b_i(n)\in\{0,1\}$, $i\in\{1, 2, 3, \ldots, m\}$ and $n \in 2^m-1$ is any entire binary vector set (AS) consisting from all possible $2^m$ binary combinations. Then, the vector space (composed of  $m$ bit vectors $A(n)$) formed according to (\ref{eqn_1}) is of dimension $m$ and consists of $2^m$ vectors, which is why the vectors $(A(n))$ can be used as ASs. For further investigations, the set of vectors $B(n)$ is regarded as linear ASs or simply binary up-counter sequences. The enormous variety of AS generated according to~(\ref{eqn_1}) primarily depends on the values of linearly independent vectors ($v_i = \beta_1(i) \beta_2(i) \ldots \beta_{m-1}(i) \beta_m(i)$, $\beta_j(i) \in \{0,1\}$, $j=\overline{1,m}$, ), which form the generating binary $m \times m$ matrix $V$. The only restriction for such a matrix ($V$) is the maximal rank achieved by choosing a linearly independent set of vectors $v_i$. The second argument extending the possibilities to generate different ASs is the vector set $B(n)$. This set consists of all possible $2^m$ binary vectors. Thus, any AS can be used as the vector set $B(n)$ for generation of new AS according to (\ref{eqn_1}).

Brief analyses of the above-presented Relation (\ref{eqn_1}), which can be used for AS generation, reveal at least two questions. The first question concerns the generation matrix $V$, and the second question addresses the computational complexity of the above-presented algorithm (\ref{eqn_1}).

\medskip
The rank of a random $m \times m$ matrix $V$ with entries in $GF(2)$, which are independently chosen and equally likely to be $0$ or $1$ ($p(0) = p(1) = 0, 5$), is analyzed in Kolchin's book~\cite{Kolchin_1998_book}. He proved that the probability that the rank of a random $m \times m$ matrix is $m - s$ equals:
\begin{equation}
\label{_eqn_2}
P(m,s)=  2^{-s^2}(\prod_{\mathclap{0\leq i<\leq m -s -1}} (1 - 2^{-(m-i)})\times
(\sum_{\mathclap{0\le i1 \le i2 \ldots is \le m-s}}2^{-i1 - i2 \ldots is}).
\end{equation}
In the case of $s = 0$, the probability of the full rank matrix is as follows:
\begin{equation}
\label{eqn_3}
P(m,0) = \prod_{i=0}^{m}(1-2^{-i}).
\end{equation}

\medskip
As $m$ becomes larger and larger, $P(m, s)$ approaches the limiting values, for example, if $s = 0$ and $m \to \infty$, then $P(m,0)~\approx~0,2887880950866$. At the same time, the expected value of the rank of the random matrix is $m-\sum_{s=0}^{m}sP(m,s)$.  For a large value of $m$, this value approaches $m-0,850179830874$~\cite{Kolchin_1998_book}. Thus, the number of ASs generated according to (\ref{eqn_1}) for real values of $m$ reaches astronomical values that are equal to more than 28.8\% of the total number $2^{m^2}$ of possible $m \times m$ binary matrices. The procedure of AS generation based on Relation (\ref{eqn_1}) for the case of $m = 4$ and
the set of $m$ linearly independent binary vectors $v_1=1011, v_2=1000, v_3=0101, v_4=1111$ forming the generated matrix $V$ is presented in Table~\ref{tab_3_procedure}:
\begin{equation}
\label{eqn_4}
V=
\begin{vmatrix}
v_1\\
v_2\\
v_3\\
v_4
\end{vmatrix}
=
\begin{vmatrix}
1 & 0 & 1 & 1 \\
1 & 0 & 0 & 0 \\
0 & 1 & 0 & 1 \\
1 & 1 & 1 & 1
\end{vmatrix}
\end{equation}

\begin{table}[!h]
\vspace*{-4mm}
	\caption{Procedure of address sequence generation based on Relation ~(\ref{eqn_1})}
	\label{tab_3_procedure}\vspace*{-1mm}
	\centering
	\resizebox{0.78\textwidth }{!}{%
		\begin{tabular}{cccc}
			\toprule
			$n$ & $B(n)=b_4(n)b_3(n)b_2(n) b_1(n)$ & $A(n)=a_4(n)a_3(n)a_2(n)a_1(n)$ &
			$B(n) \oplus B(n-1)$   \\
			\midrule
			$
			\begin{matrix}
			0 \\
			1 \\
			2 \\
			3 \\
			4 \\
			5 \\
			6 \\
			7 \\
			8 \\
			9 \\
			10 \\
			11 \\
			12 \\
			13 \\
			14 \\
			15 \\
			\end{matrix}
			$
			&
			$\begin{matrix}
			0 &   0 &   0 &   0 \\
			0 &   0 &   0 &   1 \\
			0 &   0 &   1 &   0 \\
			0 &   0 &   1 &   1 \\
			0 &   1 &   0 &   0 \\
			0 &   1 &   0 &   1 \\
			0 &   1 &   1 &   0 \\
			0 &   1 &   1 &   1 \\
			1 &   0 &   0 &   0 \\
			1 &   0 &   0 &   1 \\
			1 &   0 &   1 &   0 \\
			1 &   0 &   1 &   1 \\
			1 &   1 &   0 &   0 \\
			1 &   1 &   0 &   1 \\
			1 &   1 &   1 &   0 \\
			1 &   1 &   1 &   1 \\
			\end{matrix}$
			&
			$
			\begin{matrix}
			0 &   0 &   0 &   0 \\
			1 &   0 &   1 &   1 \\
			1 &   0 &   0 &   0 \\
			0 &   0 &   1 &   1 \\
			0 &   1 &   0 &   1 \\
			1 &   1 &   1 &   0 \\
			1 &   1 &   0 &   1 \\
			0 &   1 &   1 &   0 \\
			1 &   1 &   1 &   1 \\
			0 &   1 &   0 &   0 \\
			0 &   1 &   1 &   1 \\
			1 &   1 &   0 &   0 \\
			1 &   0 &   1 &   0 \\
			0 &   0 &   0 &   1 \\
			0 &   0 &   1 &   0 \\
			1 &   0 &   0 &   1 \\
			\end{matrix}
			$
			&
			$\begin{matrix}
			1 &   1 &   1 &   1 \\
			0 &   0 &   0 &   1 \\
			0 &   0 &   1 &   1 \\
			0 &   0 &   0 &   1 \\
			0 &   1 &   1 &   1 \\
			0 &   0 &   0 &   1 \\
			0 &   0 &   1 &   1 \\
			0 &   0 &   0 &   1 \\
			1 &   1 &   1 &   1 \\
			0 &   0 &   0 &   1 \\
			0 &   0 &   1 &   1 \\
			0 &   0 &   0 &   1 \\
			0 &   1 &   1 &   1 \\
			0 &   0 &   0 &   1 \\
			0 &   0 &   1 &   1 \\
			0 &   0 &   0 &   1 \\
			\end{matrix}
			$ \\
			\bottomrule
	\end{tabular}
}
\end{table}

\medskip
\noindent In this case:
\begin{equation*}
A(n) =b_1(n) \times v_1  \oplus b_2(n) \times v_2   \oplus b_3(n) \times v_3    \oplus  b_4(n) \times v_4.
\end{equation*}
For example $A(5) =b_1(5) \times v_1   \oplus b_2(5) \times v_2  \oplus b_3(5) \times v_3  \oplus  b_4(5) v_4   = 1 \times v_1 \oplus 0 \times v_2 \oplus 1 \times v_3 \oplus 0 \times v_4$ and finally we have  $A(5)= v_1 \oplus v_3 = 1011 \oplus 0101 = 1110$.

\medskip
Table~\ref{tab_3_procedure} reveals that the number of operands for consecutive address $A(n)$ calculation according to~(\ref{eqn_1}) strongly depends on the number of $1$s within the $B(n)$. Depending on the number of nonzero components of $B(n)$, up to $m$ operands can obtain the value of $A(n)$. This indicates that the address generation according to~(\ref{eqn_1}) is a time-consuming procedure, which may sufficiently reduce the rate of the test pattern generation.

\medskip
To decrease the number of operations to only one bitwise XOR operation, Relation (\ref{eqn_1}) can be transformed into a recursive relation~\cite{Chen_2007_562_568, Yarmolik_2013_242_247}:
\begin{equation}
\label{eqn_5}
A(n)= A(n-1) \oplus v_{i}^{*};\, n=\overline{0,2^m - 1},\, i=\overline{1,m}
\end{equation}
where
\begin{equation}
\label{eqn_6}
v_{i}^{*} = v_1 \oplus v_2 \oplus \ldots \oplus v_i.
\end{equation}

The main idea behind this transformation is based on the set of consecutive values of $B(n) \oplus B(n-1)$. Table~\ref{tab_3_procedure} indicates that only the four different correction values, $v_{1}^{*},v_{2}^{*}, v_{3}^{*}, v_{4}^{*} $, are used to obtain $A(n)$ from the previous value $A(n-1)$. For our previous example in Table~\ref{tab_3_procedure},
$v_{1}^{*} = v_{1}=1011; v_{2}^{*} = v_{1} \oplus v_{2} = 0011;
v_{3}^{*} = v_{1} \oplus v_{2} \oplus v_{3} = 0110;
v_{4}^{*} = v_{1} \oplus v_{2} \oplus v_{3} \oplus v_{4}=1001 $.

\medskip
For the general case, the recursive relation in (\ref{eqn_5}) for the AS generation can be obtained from (\ref{eqn_1}) using the new bases $V^*$ constructed according to generation matrix $V$~(\ref{eqn_6}), which is built from the linearly independent vectors, $v_i = \beta_1(i) \beta_2(i) \ldots \beta_{m-1}(i) \beta_m(i), \beta_j(i)\in \{0,1\},j=\overline{1,m}$. The same relation in (\ref{eqn_6}) can be used to obtain  $V$ from $V^*$, namely, $v_i= v_{i-1}^{*} \oplus v_{i}^{*}$. The value of the index $i$ of the binary vector $v_{i}^{*}$, which is used as a term in Expression~(\ref{eqn_5}), depends on the so-called switching sequence,~$T_m$, of the reflected gray code~\cite{Savage_996_605_629}. The binary reflected gray code, also known as the standard gray code, is the best-known gray code~\cite{Savage_996_605_629}. A characteristic property of the binary standard gray code is that the second half of the list of codewords can be obtained from the first half by reflection (i.e., by writing the first half backwards and replacing the first $0$ with $1$). Any reflected gray code is described by the switching sequence~$T_m$. For example, for $m = 4$, this sequence has the form $T_4 = 1, 2, 1, 3, 1, 2, 1, 4, 1, 2, 1, 3, 1, 2, 1$. Formally, the switching sequence $T_m$ defines the index $i$ of an inverted bit to obtain the new value $B(n)_g$ from the previous value $B(n-1)_g$. Index $g$ of the $B(n)_g$ indicates the representation in the gray code of the initial binary code $B(n) = b_m(n)b_{m-1}(n)b_{m-2}(n) \ldots b_2(n)b_1(n)$. The vector $B(n)_g$ in the gray code $B(n)_g = g_m(n)\allowbreak g_{m-1}(n) \allowbreak g_{m-2}(n) \ldots g_2(n) g_1(n)$ can be obtained according to the following well-known relation~\cite{Savage_996_605_629}:
\begin{equation}
\label{eqn_7}
\begin{aligned}
& g_m(n) = b_m(n)\\
& g_i(n) = b_{i+1}(n) \oplus b_i(n) ; i=\overline{1,m-1}.\\
\end{aligned}
\end{equation}

The values of the bits of gray code $B(n)_g$ for $m = 4$ are determined in accordance with the relations $g_4(n) = b_4(n)$, $g_3(n) = b_4(n) \oplus b_3(n)$, $g_2(n) = b_3(n) \oplus b_2(n)$, and $g_1(n) = b_2(n) \oplus b_1(n)$.

\section{Address Sequence Generator}

\begin{figure}[!t]
\vspace*{-1mm}
	\centering
	\includegraphics[width=0.7\textwidth]{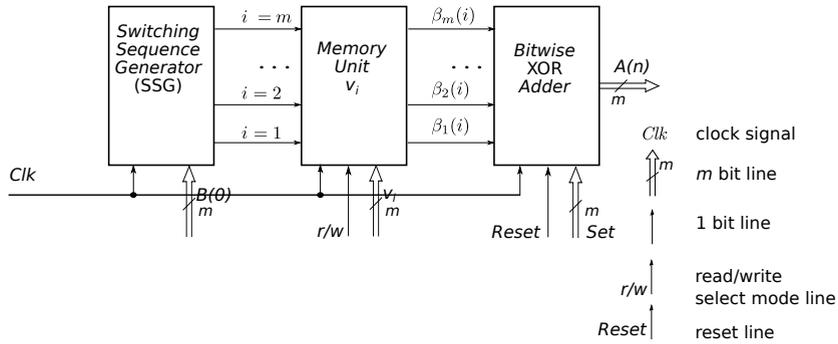}\vspace*{-1.6mm}
	\caption{General structure of the proposed address sequence generator.}
	\label{fig_1}\vspace*{-2mm}
\end{figure}

The general structure of the proposed ASG consists of three sequentially connected function blocks, as presented in Fig.~\ref{fig_1}. The first block, the switching sequence generator (SSG), is used to select one out of $m$ vectors $v_i$ per clock ({\em Clk}) according to the required order. As demonstrated in the previous section, the main block of the ASG is a memory block for storing $m$ linearly independent vectors $v_i = \beta_1(i) \beta_2(i) \ldots \beta_{m-1}(i) \beta_m(i), \beta_j(i)\in \{0,1\},j=\overline{1,m}$, which form the binary $m \times m$ generation matrix~$V$. Then, the second block is the {\em memory unit} consisting of $m$ $m$-bit cells for storing vectors~$v_i$. Memory units must perform {\em read} ({\em r}) and {\em write} ({\em w}) operations ({\em r/w}) for the generation of vectors $v_i$ and upload the new values (Fig.~\ref{fig_1}). The last block is the {\em bitwise} XOR {\em adder} to perform the operation $A(n) = A(n-1) \oplus v_i$. It consists of $m$ synchronous $D$-type flip-flops and $m$ two-input XOR gates. The adder moves to the next state after the clock pulse (\textit{Clk}) generation. Set and reset inputs in \textit{D}-type flip-flops load the all-zero state $ A(0) $ (\textit{reset}) to the adder or any other initial state $ A(0) $ (\textit{set}). At the output of the adder, the desired AS is generated.

\begin{figure}[!h]
\vspace*{1mm}
	\centering
	\includegraphics[width=0.64\textwidth]{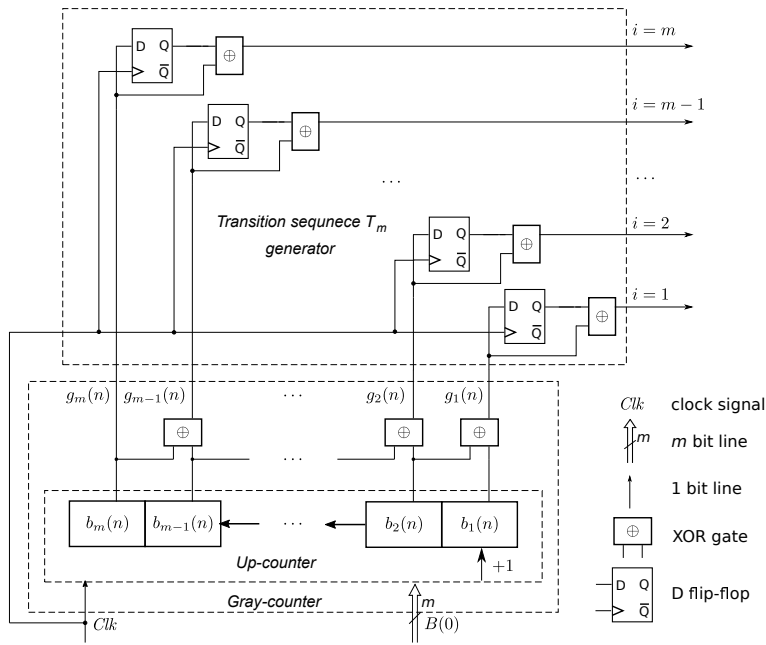}\vspace*{-1.6mm}
	\caption{Switching sequence $T_m$ generator.}
	\label{fig_2}
\end{figure}

The structure of the second (memory unit) and third (bitwise XOR adder) blocks is quite simple and standard, but the construction of the first block is not as obvious. The architecture of this block, namely, the SSG proposed in this paper, is characterized in Fig.~\ref{fig_2}.

The main block of the ASG is the SSG, $T_m$, which determines the sequence of the selection of the vectors $ v_i $ of the matrix $ V $ (~\ref{fig_2}). As noted, in each cycle of operation, only one output of the SSG generates an enable signal that determines the selected vector $ v_i $ by its index $ i $.

The \textit{up-counter} is the main part of the SSG, which consists of the $ m $-bit binary counter that performs two micro-operations. The first operation is the \textit{increment by} 1 (+1) operation to make the transition from $B(n-1)$ to $B(n)$. The second operation is the \textit{load} instruction, which is used to upload the initial state $ B(0) $ to the up-counter. Both operations are synchronous and are performed by the clock signal \textit{Clk}. The \textit{gray counter} in Fig.~\ref{fig_2} consists of an up-counter and $ m $–1 two-input XOR gates connected according to~(\ref{eqn_7}). The SSG $ T_m $ performs the bitwise XOR operation between consecutive vectors $B(n-1)_g$ and $B(n)_g$  to obtain one out of $m$ output selection vector $v_i$ signals. It consists of $m$ $D$-type flip-flops and $m$ two-input XOR gates.

\medskip
The proposed UASG illustrated in Figs.~\ref{fig_1}~and~\ref{fig_2} generates any AS depending on the values of the generated matrix $V$. The only restriction for $V$ is the linear independence of the binary vectors $ v_i $. For example, for $ m = 4 $ and the matrix~(\ref{eqn_4}) the procedure for obtaining the AS (set of all possible $m$ bit vectors $A(n)$)  in detail is presented in Table~\ref{tab_4_procedure_of_as_generation}.

\begin{table}[!h]
\vspace*{-2mm}
	\caption{Procedure of AS generation based on relation (\ref{eqn_5})}
	\label{tab_4_procedure_of_as_generation}\vspace*{-1mm}
	\centering
	\resizebox{0.79\textwidth }{!}{%
		\begin{tabular}{c|c|c|c|c|c|c|c|c}
			\toprule
			$n$ & $B(n)$ & $B(n)_g$ & $B(n)_g \oplus B(n-1)_g$ & $T_m$ & $v_{i}$ & $\Uparrow A(n)$ & $\Downarrow A(n)$ & $\Uparrow A^{*}(n)$   \\
			\midrule
			$
			\begin{matrix}
			0 \\
			1 \\
			2 \\
			3 \\
			4 \\
			5 \\
			6 \\
			7 \\
			8 \\
			9 \\
			10 \\
			11 \\
			12 \\
			13 \\
			14 \\
			15 \\
			\end{matrix}
			$
			&
			$\begin{matrix}
			0 &   0 &   0 &   0 \\
			0 &   0 &   0 &   1 \\
			0 &   0 &   1 &   0 \\
			0 &   0 &   1 &   1 \\
			0 &   1 &   0 &   0 \\
			0 &   1 &   0 &   1 \\
			0 &   1 &   1 &   0 \\
			0 &   1 &   1 &   1 \\
			1 &   0 &   0 &   0 \\
			1 &   0 &   0 &   1 \\
			1 &   0 &   1 &   0 \\
			1 &   0 &   1 &   1 \\
			1 &   1 &   0 &   0 \\
			1 &   1 &   0 &   1 \\
			1 &   1 &   1 &   0 \\
			1 &   1 &   1 &   1 \\
			\end{matrix}$
			&
			$
			\begin{matrix}
			0 &   0 &   0 &   0 \\
			0 &   0 &   0 &   1 \\
			0 &   0 &   1 &   1 \\
			0 &   0 &   1 &   0 \\
			0 &   1 &   1 &   0 \\
			0 &   1 &   1 &   1 \\
			0 &   1 &   0 &   1 \\
			0 &   1 &   0 &   0 \\
			1 &   1 &   0 &   0 \\
			1 &   1 &   0 &   1 \\
			1 &   1 &   1 &   1 \\
			1 &   1 &   1 &   0 \\
			1 &   0 &   1 &   0 \\
			1 &   0 &   1 &   1 \\
			1 &   0 &   0 &   1 \\
			1 &   0 &   0 &   0 \\
			\end{matrix}
			$
			&
			$\begin{matrix}
			0 &   0 &   0 &   0 \\
			0 &   0 &   0 &   1 \\
			0 &   0 &   1 &   0 \\
			0 &   0 &   0 &   1 \\
			0 &   1 &   0 &   0 \\
			0 &   0 &   0 &   1 \\
			0 &   0 &   1 &   0 \\
			0 &   0 &   0 &   1 \\
			1 &   0 &   0 &   0 \\
			0 &   0 &   0 &   1 \\
			0 &   0 &   1 &   0 \\
			0 &   0 &   0 &   1 \\
			0 &   1 &   0 &   0 \\
			0 &   0 &   0 &   1 \\
			0 &   0 &   1 &   0 \\
			0 &   0 &   0 &   1 \\
			\end{matrix}
			$
			&
			$
			\begin{matrix}
			4 \\
			1 \\
			2 \\
			1 \\
			3 \\
			1 \\
			2 \\
			1 \\
			4 \\
			1 \\
			2 \\
			1 \\
			3 \\
			1 \\
			2 \\
			1 \\
			\end{matrix}
			$
			&
			$
			\begin{matrix}
			1 & 1 & 1 & 1 \\ 
			1 & 0 & 1 & 1 \\ 
			1 & 0 & 0 & 0 \\ 
			1 & 0 & 1 & 1 \\ 
			0 & 1 & 0 & 1 \\ 
			1 & 0 & 1 & 1\\ 
			1 & 0 & 0 & 0 \\ 
			1 & 0 & 1 & 1 \\ 
			1 & 1 & 1 & 1 \\ 
			1 & 0 & 1 & 1 \\ 
			1 & 0 & 0 & 0 \\ 
			1 & 0 & 1 & 1 \\ 
			0 & 1 & 0 & 1 \\ 
			1 & 0 & 1 & 1\\ 
			1 & 0 & 0 & 0 \\ 
			1 & 0 & 1 & 1\\ 
			\end{matrix}
			$
			&
			$\begin{matrix}
			0 &   0 &   0 &   0 \\ 
			1 &   0 &   1 &   1 \\ 
			0 &   0 &   1 &   1 \\ 
			1 &   0 &   0 &   0 \\ 
			1 &   1 &   0 &   1 \\ 
			0 &   1 &   1 &   0 \\ 
			1 &   1 &   1 &   0 \\ 
			0 &   1 &   0 &   1 \\ 
			1 &   0 &   1 &   0\\ 
			0 &   0 &   0 &   1 \\ 
			1 &   0 &   0 &   1 \\ 
			0 &   0 &   1 &   0 \\ 
			0 &   1 &   1 &   1 \\ 
			1 &   1 &   0 &   0 \\ 
			0 &   1 &   0 &   0 \\ 
			1 &   1 &   1 &   1 \\ 
			\end{matrix}$
			&
			$
			\begin{matrix}
			1 &  1 &  1 &  1 \\ 
			0 &  1 &  0 &  0 \\ 
			1 &  1 &  0 &  0 \\ 
			0 &  1 &  1 &  1 \\ 
			0 &  0 &  1 &  0 \\ 
			1 &  0 &  0 &  1 \\ 
			0 &  0 &  0 &  1 \\ 
			1 &  0 &  1 &  0 \\ 
			0 &  1 &  0 &  1 \\ 
			1 &  1 &  1 &  0 \\ 
			0 &  1 &  1 &  0 \\ 
			1 &  1 &  0 &  1 \\ 
			1 &  0 &  0 &  0 \\ 
			0 &  0 &  1 &  1 \\ 
			1 &  0 &  1 &  1 \\ 
			0 &  0 &  0 &  0 \\ 
			\end{matrix}
			$
			&
			$\begin{matrix}
			1 &  0 &  0 &  0 \\
			0 &  0 &  1 &  1 \\
			1 &  0 &  1 &  1 \\
			0 &  0 &  0 &  0 \\
			0 &  1 &  0 &  1 \\
			1 &  1 &  1 &  0 \\
			0 &  1 &  1 &  0 \\
			1 &  1 &  0 &  1 \\
			0 &  0 &  1 &  0 \\
			1 &  0 &  0 &  1 \\
			0 &  0 &  0 &  1 \\
			1 &  0 &  1 &  0 \\
			1 &  1 &  1 &  1  \\
			0 &  1 &  0 &  0 \\
			1 &  1 &  0 &  0 \\
			0 &  1 &  1 &  1 \\
			\end{matrix}
			$
			\\
			\bottomrule
	\end{tabular}\vspace*{1mm}
}
\end{table}

The first column contains the binary values $B(n) = b_4(n)b_3(n)b_2(n)b_1(n)$  of the \textit{up-counter} starting from the all-zero state. Transformed into gray code, $B(n)_g = g_4(n)g_3(n)g_2(n)g_1(n)$ vectors as the output sequence of the \textit{gray counter} are listed in the next column (Fig~\ref{fig_2} and Table~\ref{tab_4_procedure_of_as_generation}). Columns $B(n)_g \oplus B(n-1)_g$, $T_m$ and $v_{i}$ contain the transition sequence output signals used for selecting one $v_i = \beta_1(i) \beta_2(i)\beta_3(i)\beta_4(i)$ out of the four vectors ($v_1$, $v_2$, $v_3$, $v_4$) of the matrix (\ref{eqn_4}) as well as corresponding $v_{i}$ vectors. The output AS  (vectors $A(n) = A(n-1)\oplus v_{i}; n = 0, 1, 2, \ldots, 2^4-1, i\in \{1, 2, 3, 4\}$ in the next column ${\Uparrow}A(n)$ can be regarded as the up-sequence. The initial value for the up-sequence generation of the all-zero vector $A(0) = 0 0 0 0$ was chosen. The corresponding down-sequence, the sequence with the reversed address order, is presented in column ${\Downarrow}A(n)$. To generate the down-sequence, the initial value of ${\Downarrow}A(0)$ must be equal to the last ${\Uparrow}A(15)$ value of the up-sequence. In this case, ${\Downarrow}A(0) = {\Uparrow}A(15) = 1 1 1 1$ (Table~\ref{tab_4_procedure_of_as_generation}).

\eject

To summarize, the initial value ${\Downarrow}A(0)$ for the down-sequence generation must be equal to the final state ${\Uparrow}A(n-1)$ of the up-sequence, which follows from the properties of the XOR operation and is formulated as Statement 1.

\begin{statement}
	\label{stat_1}
	A decreasing sequence (down-sequence) of addresses ${\Downarrow}A(n), n\in \{0, 1, 2, \ldots, 2^m – 1\}$ with respect to the increasing sequence (up-sequence) of addresses ${\Uparrow}A(n), n\in \{0, 1, 2, \ldots, 2^m – 1\}$, for which ${\Downarrow}A(n) = {\Uparrow}A(2^m–1–n)$, is generated using Relation (\ref{eqn_5}) and the same generator matrix $V$ as for generating ${\Uparrow}A(n)$ with the starting address ${\Downarrow}A(0)$ equal to ${\Uparrow}A(2^m – 1)$.
\end{statement}	

If the initial state $A(0)$ is not an all-zero state, the AS  differs from the original AS obtained for the zero starting state. All bits of vector $A(n) = a_m(n)a_{m–1}(n)a_{m–2}(n)\ldots a_2(n)a_1(n)$ for which $a_j(0) = 1, j\in \{1, 2, 3, \ldots , m\}$ are inverted. Table~\ref{tab_4_procedure_of_as_generation} reveals the sequence ${\Uparrow}A^*(n)$ for which $A^*(0) = 1 0 0 0$ is the copy of ${\Uparrow}A(n)$ with just the fourth bit inverted. The sequence ${\Downarrow}A(n)$ was obtained from the inverted values of ${\Uparrow}A(n)$ because ${\Downarrow}A(0) = 1 1 1 1$.

\section{Address sequences}

The declared goal of the presented research is a UASG with a wide spectrum of generated sequences. The general expression for AS generation according to the algorithm implemented as the UASG is based on the recursive relation in (\ref{eqn_5}) and has the following form:
\begin{equation}
\label{eqn_8}
\begin{aligned}
& A(0) =  A;\\
& A(n) =  A(n-1) \oplus v_{i(T_m(B))} \\
&         \qquad\qquad\qquad\qquad n=\overline{1,2^m-1};  i=\overline{1,m}\\
\end{aligned}
\end{equation}

The initial conditions of UASG depend on the values of two constants $A = a_ma_{m-1}\ldots a_3a_2a_1$ and $B=b_mb_{m-1}\ldots b_3b_2b_1$, where $a_i, b_i, \in \{0, 1\}$, and $m$ $m$-bit binary vectors $v_i = \beta{1}(i) \beta_2(i)\allowbreak \ldots \beta_{m-1}(i) \beta_m(i), \beta_j(i)\in\{0,1\},j=\overline{1,m}$,  which form the binary $m \times m$ generation matrix $V$. Constants $A$ and $B$ are represented by the initial states of \textit{bitwise} XOR \textit{adder} and \textit{up-counter}, respectively (Figs.~\ref{fig_1}~and~\ref{fig_2}). Concerning the constants $A$ and $B$, no restrictions exist for their values and, usually, their standard meaning is all-zero values. For some UASG implementations, the zero initial conditions for $A$ and $B$ can save the required area for the fabrication of the generator.

The only requirement for the generation matrix $V$ is its maximal rank, which generates cyclic ASs with a length of $2^m$~\cite{Boyd_2018_book}. The required order of the vectors $v_{i}(T_m(B))$ depends on the switching sequence $T_m(B)$ of the reflected gray code generated by the \textit{gray counter} (Fig.~\ref{fig_2}). The gray counter is constructed on the bases of the binary \textit{up-counter}, which can use any initial state $ B(0) $. Usually, the starting value of the \textit{up-counter} is a zero binary vector, and the sequence of the used vectors $ v_i $ corresponds to the standard reflected gray code sequence $ T_m $. Applying nonzero values of $B(0) \ne 0 0 0 \ldots 0$ initiates the generation process of the shifted version of $ T_m = T_m(B) $, where $ B $ determines the number of shifts. Then, $i = i(T_m(B))$ is the function of $ T_m(B) $, which defines the sequence of the selected vectors  for AS generation.

The analysis of the existing memory tests reveals that it is necessary to generate addresses in the reverse sequence ${\Downarrow}A(n)$ concerning the original ${\Uparrow}A(n)$ ASs and their various modifications. To solve this problem, the peculiar properties of the proposed model (\ref{eqn_8}) for AS generation can be used for further investigation. The equivalence of the addition (XOR) and subtraction operations by modulo two ($\oplus$)~\cite{Boyd_2018_book} and the symmetry of the switching sequence $T_m$ of the reflected gray code~\cite{Savage_996_605_629} are basic features of (\ref{eqn_8}). Within the framework of the proposed model in (\ref{eqn_8}) of generating ASs, the formation of a sequence of decreasing ${\Downarrow}A(n)$ addresses concerning ${\Uparrow}A(n)$ is consistent with Statement~\ref{stat_1}~\cite{Yarmolik_V_N_2014_124_136}. Statement~\ref{stat_1} is true for any AS generated according to ((\ref{eqn_8})) and the arbitrary initial value of $ B(0) $. For all-zero values~of~$ B(0) $, the example of up and down ASs is given in Table~\ref{tab_4_procedure_of_as_generation}.

\begin{statement}
	\label{stat_2}
	A shifted copy of the AS  by any $ l  $ number of positions compared with the original AS for the case of all-zero values of $ B(0) $ and $ A(0)  $ generated according to~(\ref{eqn_8}) is obtained when $ B(0) = l $ and $ A(0) = A(l) $.
\end{statement}

Statement 2 also is true for any AS generated based on the chosen mathematical model (\ref{eqn_8}) and generates the shifted version of the address order, which is very important for multirun memory testing~\cite{Yarmolik_2007_688_698, Mrozek_2016_395_403}. As an example, the shifted copy of AS by $ l = 3 $ positions is presented in Table~\ref{tab_5_procedure_of_shifted}. For this case, $ m = 4 $ and the same matrix $ V $ (\ref{eqn_4}) is used.

\begin{table}[!h]
\vspace*{-3mm}
	\centering
	\caption{Procedure of shifted AS generation by UASG according to relation  (\ref{eqn_8})}
	\label{tab_5_procedure_of_shifted}\vspace*{-1mm}
	\resizebox{0.78\textwidth }{!}{%
		\begin{tabular}{c|c|c|c|c|c|c|c|c}
			\toprule
			\multirow{3}{0em}{$n$}
			& \multicolumn{3}{c|}{$A(n)$ with $B(0) = 0000$, } & \multicolumn{3}{c|}{$A(n)$ with $B(0) = 0011$,} & \multicolumn{2}{c}{$A(n)$ with $B(0) = 0011$,}\\
			& \multicolumn{3}{c|}{$A(0) = 0000$} & \multicolumn{3}{c|}{$A(0) = 0000$} & \multicolumn{2}{c}{$A(0) = 1000$}\\
			\cline{2-9}
			& $B(n)$ & $T_m(B)$ & $A(n)$ & $B(n)$ & $T_m(B)$ & $A(n)$ & $T_m(B)$ & $A(n)$   \\
			\midrule
			$
			\begin{matrix}
			0 \\
			1 \\
			2 \\
			3 \\
			4 \\
			5 \\
			6 \\
			7 \\
			8 \\
			9 \\
			10 \\
			11 \\
			12 \\
			13 \\
			14 \\
			15 \\
			\end{matrix}
			$
			&
			$\begin{matrix}
			0 &   0 &   0 &   0 \\
			0 &   0 &   0 &   1 \\
			0 &   0 &   1 &   0 \\
			0 &   0 &   1 &   1 \\
			0 &   1 &   0 &   0 \\
			0 &   1 &   0 &   1 \\
			0 &   1 &   1 &   0 \\
			0 &   1 &   1 &   1 \\
			1 &   0 &   0 &   0 \\
			1 &   0 &   0 &   1 \\
			1 &   0 &   1 &   0 \\
			1 &   0 &   1 &   1 \\
			1 &   1 &   0 &   0 \\
			1 &   1 &   0 &   1 \\
			1 &   1 &   1 &   0 \\
			1 &   1 &   1 &   1 \\
			\end{matrix}$
			&
			$
			\begin{matrix}
			i=4 \\
			i=1 \\
			i=2 \\
			i=1 \\
			i=3 \\
			i=1 \\
			i=2 \\
			i=1 \\
			i=4 \\
			i=1 \\
			i=2 \\
			i=1 \\
			i=3 \\
			i=1 \\
			i=2 \\
			i=1 \\
			\end{matrix}
			$
			&
			$
			\begin{matrix}
			0 & 0 & 0 & 0 \\
			1 & 0 & 1 & 1 \\
			0 & 0 & 1 & 1 \\
			1 & 0 & 0 & 0 \\
			1 & 1 & 0 & 1 \\
			0 & 1 & 1 & 0 \\
			1 & 1 & 1 & 0 \\
			0 & 1 & 0 & 1 \\
			1 & 0 & 1 & 0 \\
			0 & 0 & 0 & 1 \\
			1 & 0 & 0 & 1 \\
			0 & 0 & 1 & 0 \\
			0 & 1 & 1 & 1 \\
			1 & 1 & 0 & 0 \\
			0 & 1 & 0 & 0 \\
			1 & 1 & 1 & 1 \\
			\end{matrix}
			$
			&
			$
			\begin{matrix}
			0 & 0 & 1 & 1 \\
			0 & 1 & 0 & 0 \\
			0 & 1 & 0 & 1 \\
			0 & 1 & 1 & 0 \\
			0 & 1 & 1 & 1 \\
			1 & 0 & 0 & 0 \\
			1 & 0 & 0 & 1 \\
			1 & 0 & 1 & 0 \\
			1 & 0 & 1 & 1 \\
			1 & 1 & 0 & 0 \\
			1 & 1 & 0 & 1 \\
			1 & 1 & 1 & 0 \\
			1 & 1 & 1 & 1 \\
			0 & 0 & 0 & 0 \\
			0 & 0 & 0 & 1 \\
			0 & 0 & 1 & 0 \\
			\end{matrix}
			$
			&
			$
			\begin{matrix}
			i=1 \\
			i=3 \\
			i=1 \\
			i=2 \\
			i=1 \\
			i=4 \\
			i=1 \\
			i=2 \\
			i=1 \\
			i=3 \\
			i=1 \\
			i=2 \\
			i=1 \\
			i=4 \\
			i=1 \\
			i=2 \\
			\end{matrix}
			$
			&
			$
			\begin{matrix}
			0 & 0 & 0 & 0 \\
			0 & 1 & 0 & 1 \\
			1 & 1 & 1 & 0 \\
			0 & 1 & 1 & 0 \\
			1 & 1 & 0 & 1 \\
			0 & 0 & 1 & 0 \\
			1 & 0 & 0 & 1 \\
			0 & 0 & 0 & 1 \\
			1 & 0 & 1 & 0 \\
			1 & 1 & 1 & 1 \\
			0 & 1 & 0 & 0 \\
			1 & 1 & 0 & 0 \\
			0 & 1 & 1 & 1 \\
			1 & 0 & 0 & 0 \\
			0 & 0 & 1 & 1 \\
			1 & 0 & 1 & 1 \\
			\end{matrix}
			$
			&
			$
			\begin{matrix}
			i=1\\
			i=3\\
			i=1\\
			i=2\\
			i=1\\
			i=4\\
			i=1\\
			i=2\\
			i=1\\
			i=3\\
			i=1\\
			i=2\\
			i=1\\
			i=4\\
			i=1\\
			i=2\\
			\end{matrix}
			$
			&
			$
			\begin{matrix}
			1 & 0 & 0 & 0\\
			1 & 1 & 0 & 1\\
			0 & 1 & 1 & 0\\
			1 & 1 & 1 & 0\\
			0 & 1 & 0 & 1\\
			1 & 0 & 1 & 0\\
			0 & 0 & 0 & 1\\
			1 & 0 & 0 & 1\\
			0 & 0 & 1 & 0\\
			0 & 1 & 1 & 1\\
			1 & 1 & 0 & 0\\
			0 & 1 & 0 & 0\\
			1 & 1 & 1 & 1\\
			0 & 0 & 0 & 0\\
			1 & 0 & 1 & 1\\
			0 & 0 & 1 & 1\\
			\end{matrix}
			$
			\\
			\bottomrule
	\end{tabular}}
\end{table}

Table~\ref{tab_5_procedure_of_shifted} reveals that the AS $ A(n) $ that is shifted by $ l =3 $ positions is generated for $B(0) = 3_{(10)} = 0011_{(2)}$ and $A(3) = 1000$. For real values of $ l $, the only problem is to determine the meaning of $ A(l) $, which requires additional calculations. These calculations are based on the following statement, which follows from the above-presented discussions concerning the binary vector space~\cite{Boyd_2018_book, Kolchin_1998_book}.

\begin{statement}
	The AS, which is generated as $m$-dimensional binary vectors according to (\ref{eqn_1}) based on generating an $m \times m$ matrix $ V $ with a maximal rank, can be obtained from the recursive relation in~(\ref{eqn_5}) with the generation matrix $ V^*  $ received from (\ref{eqn_6}) and vice versa.
\end{statement}

\medskip
From this statement, it follows that, if the AS $ A(n) $ that is presented in Table~\ref{tab_5_procedure_of_shifted} is generated according to Relation (\ref{eqn_5}) or (\ref{eqn_8}) using the matrix in (\ref{eqn_4}), then the same sequence $ A(n) $ is generated based on (\ref{eqn_1}) for the generation matrix $ V^* $, which is calculated as follows:
\begin{equation}
\label{eqn_9}
V^*=
\begin{vmatrix}
v_{1}^{*} \\
v_{2}^{*} \\
v_{3}^{*} \\
v_{4}^{*} \\
\end{vmatrix}
=
\begin{vmatrix}
v_1 \\
v_1 \oplus v_2 \\
v_2 \oplus v_3 \\
v_3 \oplus v_4 \\
\end{vmatrix}
=
\begin{vmatrix}
1 & 0 & 1 & 1 \\
0 & 0 & 1 & 1 \\
1 & 1 & 0 & 1 \\
1 & 0 & 1 & 0 \\
\end{vmatrix}
\end{equation}
Applying Relation (\ref{eqn_1}) and using the matrix in (\ref{eqn_9}), $A(3) = v_{1}^* \oplus v_{2}^*  = 1 0 1 1 \oplus 0 0 1 1 = 1 0 0 0$.

\medskip
Applying different values of $ B(0) $ and $ A(0) $ generates a wide spectrum of different ASs. Among which, sequences of addresses $A(n) = a_m(n)a_{m–1}(n)a_{m–2}(n)\ldots a_2(n)a_1(n)$ with inverted bits $a_i(n), \allowbreak i\in \{1, 2 ,3, \ldots, m\}$ have actively been used in practice~\cite{Yarmolik_2007_688_698, Yarmolik_S_V_2006_212_216, Yarmolik_S_V_2006_572_576}. Considering that the bit inversion $\overline{a}_i(n)$ is equivalent to the XOR operation $\overline{a}_i(n)=a_i(n) \oplus 1$, the desired set of inverted bits within $ A(n) $ can be specified by the initial value $ A(0) $. The examples of such ASs are illustrated in Table~\ref{tab_5_procedure_of_shifted}. This technique, based on just setting a nonzero initial value $ A(0) $, allows to generate $ 2^m – 1 $ different sequences, where $m$ is the dimension of the vector space described by the generation matrix $V$.

\section{Most common Address Sequences for memory built-in self-test }

The generalized mathematical model (\ref{eqn_8}) presented in the previous section is an extension of the mathematical model used for binary vector generation. The basis of this model is in the form of the generation matrix $ V $, which determines the main properties of ASs and identifies their subsets. For memory testing, the address generator must generate several ASs because each sequence and the combinations of them have their properties that are closely related to the memory-test fault-detection capability~\cite{Goor_2011_1_6, Mrozek_2016_23_43, Mrozek_2017_1_17}. The generation of the most important and quite common ASs listed in~\cite{Goor_2011_1_6} based on UASG is considered next.

\textit{Linear} ASs, also called \textit{counting} ASs are the first in the set of an AS family. For the formation of \textit{counting} (\textit{counter}) sequences formed by binary counting circuits (counters), it is necessary to form a generation matrix $ V $ following Statement~\ref{stat_4}.

\begin{statement}
	\label{stat_4}
	The \textit{linear}(\textit{counting}) AS is generated by UASG~(\ref{eqn_8}) when the generation matrix $ V $ is the lower triangular matrix relative to the antidiagonal with only nonzero ($1s$) entries on and below the antidiagonal
\end{statement}

An example of such an AS is presented in Table~\ref{tab_6_most_used_address} for the case of $m = 4$ and $B(0) = A(0) = 0 0 0 0$. The linear AS belongs to the set of $2^j$ ASs, which generates all address pairs with a Hamming distance equal to 1~\cite{Goor_2011_1_6}. The linear AS is the $2^j$ AS with $j = 0$, obtaining the address order incremented by~1. The complete set of $2^j$ AS where $j \in \{0, 1, 2, \ldots, m-1\}$, can be generated based on the proposed solution according to the next statement.

\begin{statement}
	\label{stat_5}
	The $ 2^j $ AS is generated using UASG~(\ref{eqn_8}) when the generation matrix $ V $ is the matrix obtained as the column permutation of the lower triangular matrix relative to the antidiagonal with only nonzero ($ 1s $) entries on and below the antidiagonal with all $ 1s $ in the $m-j$ column.
\end{statement}

Table~\ref{tab_6_most_used_address} contains an example of $2^j = 4$ AS for $j = 2$, generating the addresses incremented by~4. The generation matrix $ V $ has all an $ 1s $ column with the index $m-j = 4 - 2 = 2$. In the previous section, the up and down (increasing/decreasing) sequence generation techniques for the case of any type of AS were presented. Applying the technique described by Statement~\ref{stat_1}, it is easy to generate  decreasing order of the $ 2^j $ AS. The example in Table~\ref{tab_6_most_used_address} repeats the example presented in~\cite{Goor_2011_1_6}, for which the switching activity of the other bits of $A(n) = a_4(n)a_3(n)a_2(n)a_1(n)$, except the $(j+1)$th bit corresponding to all the 1s $ m - j $ column, is constant. The proposed solutions implemented as the UASG set any switching activity for all bits of $ A(n) $, starting from the minimal $2^0 =1$ up to the maximal $2^{m-1}$~\cite{Yarmolik_2013_242_247}.

\begin{table}[!h]
\vspace*{-2mm}
	\caption{Most-used address sequence generation using the universal address sequence generator }
	\centering
	\label{tab_6_most_used_address}\vspace*{-1mm}
	\resizebox{0.75\textwidth }{!}{%
		\begin{tabular}{|c|c|c|c|c|c|c|}
			\toprule
			\multirow{2}{*}[-1.5em]{$n$} & Linear & $2^j=4$ & Complement & Limited & Gray Code & Random \\
			&
			$\begin{vmatrix}
			0 & 0 & 0 & 1 \\
			0 & 0 & 1 & 1 \\
			0 & 1 & 1 & 1 \\
			1 & 1 & 1 & 1 \\
			\end{vmatrix}
			$
			&
			$\begin{vmatrix}
			0 & 1 & 0 & 0 \\
			1 & 1 & 0 & 0 \\
			1 & 1 & 0 & 1 \\
			1 & 1 & 1 & 1 \\
			\end{vmatrix}
			$
			&
			$\begin{vmatrix}
			1 & 1 & 1 & 1 \\
			1 & 1 & 1 & 0 \\
			1 & 1 & 0 & 0 \\
			1 & 0 & 0 & 0 \\
			\end{vmatrix}
			$
			&
			$\begin{vmatrix}
			1 & 1 & 1 & 1 \\
			1 & 1 & 1 & 0 \\
			1 & 1 & 0 & 1 \\
			1 & 0 & 1 & 1 \\
			\end{vmatrix}
			$
			&
			$\begin{vmatrix}
			0 & 0 & 0 & 1 \\
			0 & 0 & 1 & 0 \\
			0 & 1 & 0 & 0 \\
			1 & 0 & 0 & 0 \\
			\end{vmatrix}
			$
			&
			$\begin{vmatrix}
			1 & 0 & 0 & 0 \\
			1 & 1 & 0 & 0 \\
			1 & 1 & 1 & 0 \\
			1 & 1 & 1 & 1 \\
			\end{vmatrix}
			$
			\\
			\midrule
			$
			\begin{matrix}
			0 \\
			1 \\
			2 \\
			3 \\
			4 \\
			5 \\
			6 \\
			7 \\
			8 \\
			9 \\
			10 \\
			11 \\
			12 \\
			13 \\
			14 \\
			15 \\
			\end{matrix}
			$
			&
			$\begin{matrix}
			0 &   0 &   0 &   0 \\
			0 &   0 &   0 &   1 \\
			0 &   0 &   1 &   0 \\
			0 &   0 &   1 &   1 \\
			0 &   1 &   0 &   0 \\
			0 &   1 &   0 &   1 \\
			0 &   1 &   1 &   0 \\
			0 &   1 &   1 &   1 \\
			1 &   0 &   0 &   0 \\
			1 &   0 &   0 &   1 \\
			1 &   0 &   1 &   0 \\
			1 &   0 &   1 &   1 \\
			1 &   1 &   0 &   0 \\
			1 &   1 &   0 &   1 \\
			1 &   1 &   1 &   0 \\
			1 &   1 &   1 &   1 \\
			\end{matrix}$
			&
			$
			\begin{matrix}
			0 & 0 & 0 & 0\\
			0 & 1 & 0 & 0\\
			1 & 0 & 0 & 0\\
			1 & 1 & 0 & 0\\
			0 & 0 & 0 & 1\\
			0 & 1 & 0 & 1\\
			1 & 0 & 0 & 1\\
			1 & 1 & 0 & 1\\
			0 & 0 & 1 & 0\\
			0 & 1 & 1 & 0\\
			1 & 0 & 1 & 0\\
			1 & 1 & 1 & 0\\
			0 & 0 & 1 & 1\\
			0 & 1 & 1 & 1\\
			1 & 0 & 1 & 1\\
			1 & 1 & 1 & 1\\
			\end{matrix}
			$
			&
			$
			\begin{matrix}
			0 & 0 & 0 & 0\\
			1 & 1 & 1 & 1\\
			0 & 0 & 0 & 1\\
			1 & 1 & 1 & 0\\
			0 & 0 & 1 & 0\\
			1 & 1 & 0 & 1\\
			0 & 0 & 1 & 1\\
			1 & 1 & 0 & 0\\
			0 & 1 & 0 & 0\\
			1 & 0 & 1 & 1\\
			0 & 1 & 0 & 1\\
			1 & 0 & 1 & 0\\
			0 & 1 & 1 & 0\\
			1 & 0 & 0 & 1\\
			0 & 1 & 1 & 1\\
			1 & 0 & 0 & 0\\
			\end{matrix}
			$
			&
			$
			\begin{matrix}
			0 & 0 & 0 & 0\\
			1 & 1 & 1 & 1\\
			0 & 0 & 0 & 1\\
			1 & 1 & 1 & 0\\
			0 & 0 & 1 & 1\\
			1 & 1 & 0 & 0\\
			0 & 0 & 1 & 0\\
			1 & 1 & 0 & 1\\
			0 & 1 & 1 & 0\\
			1 & 0 & 0 & 1\\
			0 & 1 & 1 & 1\\
			1 & 0 & 0 & 0\\
			0 & 1 & 0 & 1\\
			1 & 0 & 1 & 0\\
			0 & 1 & 0 & 0\\
			1 & 0 & 1 & 1\\
			
			\end{matrix}
			$
			&
			$
			\begin{matrix}
			0 & 0 & 0 & 0\\
			0 & 0 & 0 & 1\\
			0 & 0 & 1 & 1\\
			0 & 0 & 1 & 0\\
			0 & 1 & 1 & 0\\
			0 & 1 & 1 & 1\\
			0 & 1 & 0 & 1\\
			0 & 1 & 0 & 0\\
			1 & 1 & 0 & 0\\
			1 & 1 & 0 & 1\\
			1 & 1 & 1 & 1\\
			1 & 1 & 1 & 0\\
			1 & 0 & 1 & 0\\
			1 & 0 & 1 & 1\\
			1 & 0 & 0 & 1\\
			1 & 0 & 0 & 0\\
			\end{matrix}
			$
			&
			$
			\begin{matrix}
			1 & 0 & 0 & 0\\
			0 & 0 & 0 & 0\\
			1 & 1 & 0 & 0\\
			0 & 1 & 0 & 0\\
			1 & 0 & 1 & 0\\
			0 & 0 & 1 & 0\\
			1 & 1 & 1 & 0\\
			0 & 1 & 1 & 0\\
			1 & 0 & 0 & 1\\
			0 & 0 & 0 & 1\\
			1 & 1 & 0 & 1\\
			0 & 1 & 0 & 1\\
			1 & 0 & 1 & 1\\
			0 & 0 & 1 & 1\\
			1 & 1 & 1 & 1\\
			0 & 1 & 1 & 1\\
			\end{matrix}
			$
			\\
			\bottomrule
		\end{tabular}
	}
\end{table}

The \textit{complement} AS described in~\cite{Goor_2011_1_6} specifies the sequence that, in the even cycle, represents the linear up-sequence and, in the odd cycle, takes the complementary value of the preceding even cycle.

\begin{statement}
	The \textit{complement} AS is generated by the UASG~(\ref{eqn_8}) when the generation matrix $ V $ is the upper triangular matrix relative to the antidiagonal with only nonzero ($1s$) entries on and above the antidiagonal.
\end{statement}

For the case of $ m = 4 $, Table~\ref{tab_6_most_used_address} contains the set of complement addresses that takes complementary values from the odd cycle of the previous address obtained from the even cycle. In the even cycle, (Table~\ref{tab_6_most_used_address}) this sequence corresponds to the linear AS. The \textit{complement} AS is extremely useful for providing the stress behavior of a memory address decoder. In this case, the high rate of switching activity of the address bits creates considerable noise, a high level of power consumption, and maximal delay~\cite{Goor_2011_1_6}.

The same high efficiency can be obtained in terms of the speed-related memory faults using an AS with \textit{limited} switching activity~\cite{Yarmolik_V_N_2014_124_136}. This type of AS obtains addresses with the highest possible switching activity. To generate such ASs, the generation matrix $ V $ should satisfy the following statement.

\begin{statement}
	\label{stat_7}
	The generation matrix $ V $ of UASG~(~\ref{eqn_8}) for the AS with \textit{limited} (highest possible) switching activity consists of one unit column and $ m – 1 $ columns that are different from each other, containing a $0$ value in each of the $ m – 1 $ rows, except the first row.
\end{statement}

 For the case of $ m = 4 $, Table~\ref{tab_6_most_used_address} contains the set of addresses with \textit{limited} activity that takes reflected grey codes in the even cycles and complementary values in the odd cycle of the previous address obtained in the even cycle (Table~\ref{tab_6_most_used_address}).  Like the \textit{complement} AS, the \textit{limited} AS is also especially useful for providing the stress behavior of the memory address decoder.

 To minimize the stress during memory testing, sequences with minimal switching activity are used, among those in the first place is the \textit{gray code} AS. In the general case, the AS with the minimum switching activity (minimum Hamming distance) formed according to~(\ref{eqn_8}) is provided by the matrix $ V $ with a minimum number of nonzero values. For an arbitrary case, such a matrix is constructed according to Statement~\ref{stat_8}.

 \begin{statement}
 	\label{stat_8}
	The generation matrix $ V $ for the AS generation based on~(\ref{eqn_8}) with minimum switching activity consists of $m$ rows that are different from each other, each of which contains just one value of~$1$.
 \end{statement}	

 According to the above statement, such a generation matrix $ V $ characterized by a set of columns differing from each other, containing one nonzero value, as listed in Table~\ref{tab_6_most_used_address} for the case of $ m = 4 $. This example is the standard reflected gray code sequence. In addition, $m!$ different gray code ASs exist, which can be reproduced by the UASG. In the case of $ m = 4 $, this number equals $ 4! = 24 $ as a result of all permutations of the matrix $ V $ columns resulting in the bit rearrangement of the address $ A(n) $.

 All the above-described ASs belong to the set of so-called deterministic sequences that are widely used for MBIST. The next widely used set of ASs for memory testing involves so-called \textit{pseudorandom} sequences that are sequences of nonrandom numbers that have properties of random sequences. The $M$--sequences generated by LFSR are often used as ASs~\cite{Goor_2011_1_6, Saravanan_2019_239_247,Pavani_2016_1484_1488, Singh_2011)561_565}. The \textit{quasi-random} sequences also belong to the family of sequences which, being deterministic, have the main properties of random sequences~\cite{Chen_2007_562_568, Yarmolik_V_N_2014_124_136, Saravanan_2019_239_247, Liu_2016_1896_1909}. To have a similar computation overhead to pseudorandom testing, quasi-random testing uses quasi-random sequences to generate low-discrepancy and low-dispersion test cases that help deliver high fault-detection effectiveness~\cite{Liu_2016_1896_1909}.

 The mathematical model described by Relation~(\ref{eqn_8}) and matrix $ V $ of directed numbers ($m$-bit binary vectors) in the form of a lower triangular matrix with a unit diagonal can be used for the case of ASs related to quasi-random sequence generation. In the general case, any square matrix $ V $ with the properties described in Statement~\ref{stat_9} can be used for quasi-random AS generation.

 \begin{statement}
 	\label{stat_9}
 	The generation matrix $ V $ for quasi-random AS generation based on~(\ref{eqn_8}) has the form of a lower triangular matrix with all $ 1s $ on the main diagonal.
 \end{statement}

The specific values of the binary vectors of the lower triangular matrix correspond to the modified directed numbers that specify a specific form of the quasi-random sequence. For example, in the case in which all entries below and on the main diagonal are $1s$, the generated AS is a \textit{van der Corput} sequence~\cite{Yarmolik_2013_242_247, Yarmolik_V_N_2014_124_136, Liu_2016_1896_1909}. For the case of $ m = 4 $, this sequence is presented in Table~\ref{tab_6_most_used_address}.

\section{Hardware implementation}

The hardware overhead of the proposed solution for the ASIC circuitry can be estimated based on the general structure of the proposed address sequence generator (ASG) in Fig.~\ref{fig_1}. The UASG consists of three main blocks as depicted in Section 3, namely the switching sequence generator (SSG), memory unit, and bitwise XOR adder. For the general case with $m$-bit addresses, the hardware overhead is needed for all these blocks, and the whole UASG is shown in the next table.

\begin{table}[!h]
\vspace*{-3mm}
	\caption{Hardware overhead for UASG implementation}
	\label{tab_5_hardware_overhead}\vspace*{-1mm}
	\centering
	\resizebox{0.78\textwidth }{!}{%
		\begin{tabular}{r|c|c|c|c}
			\midrule
			Standard elements & D-type flip-flops & 2 XOR gates & Memory & Up-Counter \\
			\midrule
			SSG	& $m$	& $2m - 1$	& -- &	$m$-bit up-counter\\
			\midrule
			Memory unit & & & $m$ $m$-bit memory cells & \\
			\midrule
			XOR adder & $m$ & $m$ & & \\
			\midrule
			\midrule
			\textbf{UASG} & $2m$ & $3m-1$ & $m$ $m$-bit memory cells & $m$-bit up-counter \\
			\midrule
		\end{tabular}
	}
\end{table}

Compared with the known solutions, especially the best solution~\cite{Goor_2011_1_6}, the proposed ASG requires moderate hardware overhead. According to the summarized data in Table~\ref{tab_5_hardware_overhead}, only roughly $3m$ D‑type flip-flops, $3m$ 2-XOR gates, and memory with $m$ cells each of $m$-bit size are needed for UASG implementations. Depending of the used technology the area overhead for UASG implementation will vary sufficiently and will be comparable with all known solutions.

\medskip
The performance of the proposed generator depends only on up-counter delays because this block is the slowest one (see Fig.~\ref{fig_2}). The signal delays on the bitwise XOR adder and transition sequence $T_m$ generator are the same and equals to the delay on D-type flip-flop and XOR gate (see Fig~\ref{fig_1} and Fig.~\ref{fig_2}). As well as the delay on the memory unit can be estimated as the delay for read operation by the so-called memory cycle. This delay can be measured by the delay on only one D-type flip-flop in a case of register type of the memory. In a worse case the total delay of UASG will includes delay on up-counter, delay on three D-type flip-flops and delay on three XOR gates. Taking into account that up-counter consists of $m$ sequentially connected D-type flip-flops its delay time will be dominated one for the real applications when $m$ usually greater than 32. As the conclusion of this discussions can be stated that frequency of formation of synchronizing signals ($Clk$) in the proposed UASG generator will be the same as in most similar generators utilized binary counter.

The best known solution allows generating only seven types of address sequences~\cite{Goor_2011_1_6}. Additional options in the existing solutions for new address sequence generation require additional \mbox{hardware}, \mbox{decreasing} its rough performance. In the case of UASG, any possible address sequence can be generated without additional hardware with the same performance.

\begin{figure}[!h]
	\centering
	\includegraphics[width=0.76\textwidth]{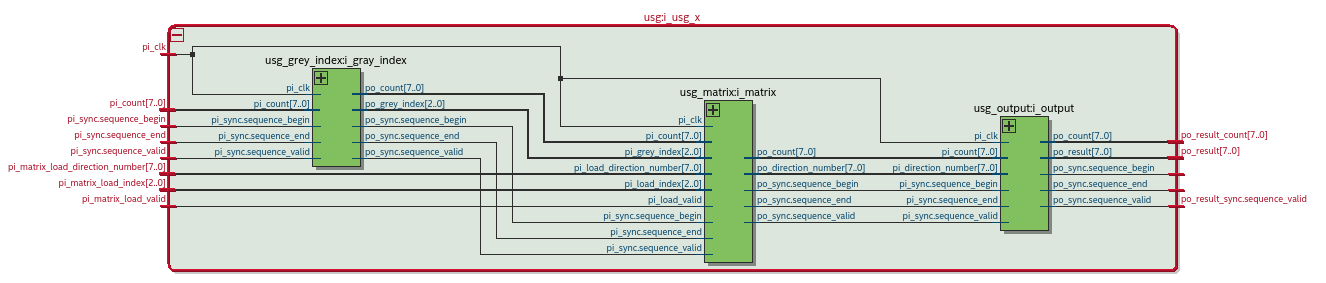}
	\caption{Implementation of an address sequence generator on FPGA.}
	\label{fig_3_implementation}
\end{figure}

\medskip
The main characteristics of the ASG were investigated using its implementation on FPGA, Intel Cyclone V (5CSXFC6D6F31C8ES), illustrated in Fig.~\ref{fig_3_implementation}. The specified FPGA consists of 41910 adaptive logic modules and 553 SRAM memory blocks (M10k). The generator implementation for $ m = 8 $ required 17 adaptive logic models and one M10k internal memory unit, which is less than 1\% of the FPGA chip area. The timing parameters of the generator correspond to the maximum possible timing parameters of the FPGA.

\begin{figure}[!h]
\vspace*{2mm}
	\centering
	\includegraphics[width=0.76\textwidth]{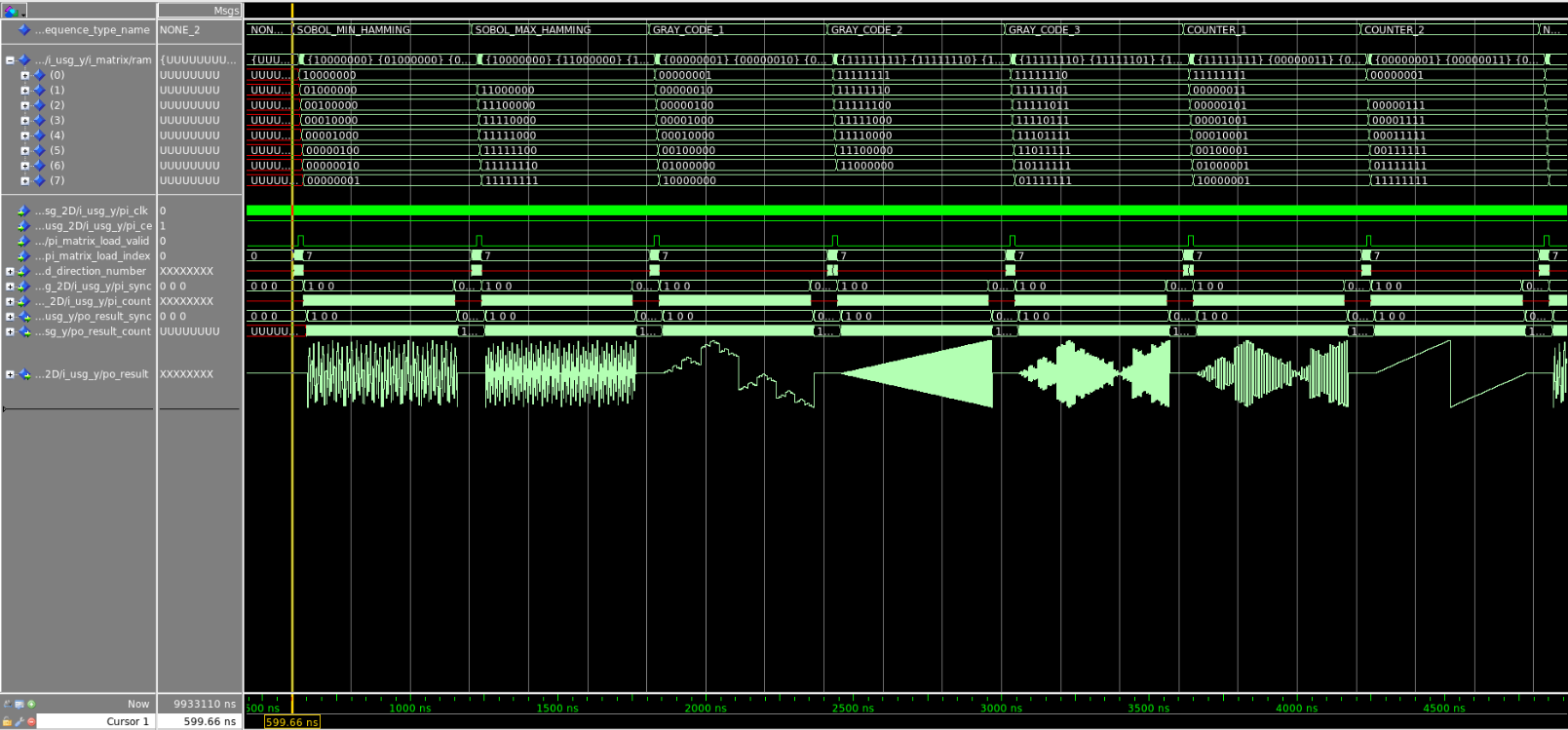}
	\caption{Waveform from the FPGA implementation of the address generator.}
	\label{fig_4_waveform}
\end{figure}

The implementation of the ASG in Fig.~\ref{fig_3_implementation} completely corresponds to the structure given earlier (Fig.~\ref{fig_1}). The input, output, and intermediate nodes of the implemented device (Fig.~\ref{fig_3_implementation}) and its detailed structure (Fig.~\ref{fig_1}) and descriptions are in full compliance. The examples of the generation of different ASs are illustrated in Fig.~\ref{fig_4_waveform}.

Figure~\ref{fig_4_waveform} displays the ModelSim simulation result waveform from the FPGA implementation of the generator for  $ m= 8 $. The top row reveals the type of sequence that is generated by the UASG. The eight rows below indicate the entries in the generation matrix $V$. The content of the matrix is loaded with the matrix load valid signal. The UASG then generates the AS according to the loaded matrix. The bottom row is a visualization of the generated sequences. The depicted waveforms are created from the values of the ASG output. The displayed sequences are (listed from left to right) the Sobol sequence with minimal Hamming distance, the Sobol sequence with the maximum Hamming distance, three types of gray code, and two types of counters, including a linear counter.

\medskip
The power consumption of UASG (Fig.~\ref{fig_3_implementation}) was analyzed using Quartus Prime (v. 19.1.0 Build 670 09/22/2019 SJ Lite Edition). The results of the analysis are given in Table~\ref{tab_7_power}, which indicates the minimum power consumption of the proposed device. The UASG time parameters correspond to the maximum possible FPGA time parameters.

\begin{table}[h]
\vspace*{-1mm}
	\caption{Power consumption analysis results}
	\label{tab_7_power}\vspace*{-1mm}
	\centering
	\resizebox{0.5\textwidth }{!}{%
		\begin{tabular}{r|r}
			\midrule
			Total Thermal Power Dissipation & 463.45 mW \\
			\midrule
			Core Dynamic Thermal Power Dissipation	& 14.63 mW \\
			\midrule
			Core Static Thermal Power Dissipation	& 415.27 mW \\
			\midrule
			I/O Thermal Power Dissipation	& 33.56 mW\\
			\midrule
		\end{tabular}
	}
\end{table}

\section{Conclusion}

The use of a modified mathematical model of quasi-random sequence generation expanded the capabilities of the ASG in terms of a significant increase in the number of types of such sequences. The essence of the method consists of the synthesis of the required generation matrix of maximum rank, providing the given values of switching activity. The limitations of the proposed technique are discussed, which are associated with the possible conflicting requirements for the values of the weights of the rows of the matrix and their linear independence. Examples of the use of such sequences for MBIST design are provided. A practical implementation of the ASG is presented, demonstrating the feasibility of such a~device with minimal hardware costs and maximum speed.


\end{document}